\documentclass[journal=nalefd,manuscript=article]{achemso}
\usepackage{chemformula} % Formula subscripts using \ch{}
\usepackage[T1]{fontenc} % Use modern font encodings
\usepackage{amsfonts}
\usepackage{bm}
\usepackage{multirow}
\usepackage{color}
\usepackage{multirow}
\usepackage{epsfig}
\usepackage{graphicx}
\usepackage[normalem]{ulem}
\usepackage{amsmath}
\usepackage{amssymb}
\usepackage{bm}
\usepackage{dcolumn}
\usepackage{braket}
\usepackage{bbold}
\usepackage{natbib}

\usepackage{array}
\usepackage{tabulary}
\usepackage{multirow}
\newcolumntype{C}[1]{>{\centering\arraybackslash}p{#1}}
\newcolumntype{L}[1]{>{\flushleft\arraybackslash}p{#1}}
\author{Yong-Kun Wang}
\affiliation{School of Physics, Northwest University, Xi'an 710127, China}
\alsoaffiliation{Shaanxi Key Laboratory for Theoretical Physics Frontiers, Xi'an 710127, China}

\author{Si Li}
\affiliation{School of Physics, Northwest University, Xi'an 710127, China}
\alsoaffiliation{Shaanxi Key Laboratory for Theoretical Physics Frontiers, Xi'an 710127, China}
\email{sili@nwu.edu.cn}

\author{Shengyuan A. Yang}
\affiliation{Research Laboratory for Quantum Materials, Department of Applied Physics, The Hong Kong Polytechnic University, Kowloon, Hong Kong, China}
\email{shengyuan.yang@polyu.edu.hk}
\title{Two-Dimensional Altermagnetic Iron Oxyhalides: Real Chern Topology and Valley-Spin-Lattice Coupling}


\begin{document}

	%%%%%%%%%%%%%%%%%%%%%%%%%%%%%%%%%%%%%%%%%%%%%%%%%%%%%%%%%%%%%%%%%%%%%
	%% The "entry" environment can be used to create an entry for the
%	\begin{tocentry}
%			\begin{center}
%	\includegraphics[width=1\textwidth]{TOC_Graphic}
%		\end{center}
	
%	\includegraphics[width=0.8\textwidth]{TOC_Graphic}
	%Some journals require a graphical entry for the Table of Contents.
	%This should be laid out ``print ready'' so that the sizing of the
	
	%
	%Inside the \texttt{tocentry} environment, the font used is Helvetica
	%8\,pt, as required by \emph{Journal of the American Chemical
	%The surrounding frame is 9\,cm by 3.5\,cm, which is the maximum
	%permitted for  \emph{Journal of the American Chemical Society}
	%graphical table of content entries. The box will not resize if the
	%content is too big: instead it will overflow the edge of the box.
	%
	%This box and the associated title will always be printed on a
	%separate page at the end of the document.
	%
%	\end{tocentry}
	
	\newpage

\begin{abstract}
	Altermagnets, a novel class of collinear magnetic materials, exhibit unique spin-split band structures, yet topological insulating states in intrinsic altermagnetic systems are rare. Here, we identify monolayer Fe$_2X_2$O ($X$ = Cl, Br, I) as a new family of 2D altermagnetic real Chern insulators. These materials display robust $d$-wave altermagnetic ordering, semiconducting band gaps, and nontrivial real Chern numbers per spin channel, yielding spin-polarized topological corner modes. They also feature spin-polarized valleys with strong altermagnetism-valley-spin-lattice coupling, enabling valley-selective excitation via linear dichroism and strain-induced valley polarization. In multiferroic Fe$_2$Cl$_2$O, magnetism coexists with ferroelasticity, and an applied strain can switch the Néel vector. These findings position 2D iron oxyhalides as a promising platform for exploring altermagnetism and magnetic topological states for spintronics and valleytronics.
	\end{abstract}

	\newpage

%\section{Introduction}
The emergence of topological insulators has sparked an extensive research field~\cite{hasan2010colloquium,qi2011topological,bansil2016colloquium}. The initially discovered
time-reversal-invariant topological insulators, such as 2D HgTe quantum wells~\cite{bernevig2006quantum,konig2007quantum} and 3D Bi$_2$Se$_3$~\cite{zhang2009topological,chen2009experimental}, feature topological gapless modes on $(d-1)$-dimensional boundaries if the system has dimension $d$. Later on, higher-order topological insulators were discovered, which have topological modes at
$(d-n)$-dimensional boundaries with $n>1$~\cite{zhang2013surface,benalcazar2017quantized,langbehn2017reflection,song2017d,benalcazar2017electric,schindler2018higher,xie2021higher}. Among them, an important class in 2D is the real Chern insulator (RCI)~\cite{zhao2017pt,sheng2019two}.
Indeed, the first discovered 2D materials with higher-order topology---graphidyne and graphyne family~\cite{sheng2019two,chen2021graphyne}---belong to RCIs.
In the bulk, RCIs are characterized by a $\mathbb{Z}_2$-valued topological invariant, known as the real Chern number. At the boundary, RCIs possess topological corner modes, which could be useful for applications, such as quantum dot devices and efficient lasing~\cite{kim2020multipolar,wu2023higher}.
So far, identified RCIs are mostly nonmagnetic materials, only very few magnetic RCIs were proposed, such as transition-metal-organic frameworks $X_3$(HITP)$_2$ ($X=$Co, Fe, Mn)~\cite{zhang2023magnetic}, Cr$_2$Se$_2$O~\cite{gong2024hidden}, and Mg(CoN)$_2$~\cite{han2025real}. It remains an on-going task to identify new magnetic RCI materials and explore the interplay between magnetism and real Chern topology.

Meanwhile, there is tremendous interest in altermagnetism~\cite{vsmejkal2022beyond,vsmejkal2022emerging,bai2024altermagnetism,fender2025altermagnetism,song2025altermagnets,hayami2019momentum,yuan2020giant,mazin2021prediction}. It refers to a class of collinear antiferromagnets, in which the two magnetic sublattices are connected not by inversion or translation, but some rotation operation. This leads to spin splitting in the electronic band structure even in the absence of spin-orbit coupling (SOC), and the splitting pattern is alternating in momentum space.
Altermagnetic materials can exhibit a range of unusual properties, such as anomalous Hall effect~\cite{Smejkal2022,feng2022anomalous}, generation of spin current~\cite{bai2022observation,karube2022observation,gonzalez2021efficient}, giant tunneling magnetoresistance~\cite{shao2021spin,vsmejkal2022giant}, and many others~\cite{ahn2019antiferromagnetism,zhu2023topological,li2023majorana,sun2023andreev,gu2025ferroelectric,duan2025antiferroelectric,vsmejkal2024altermagnetic,Zhang2024Finite,lin2025coulomb,antonenko2025mirror,wan2025interplay,Fan2025}. Experimentally, altermagnetism has been identified in several 3D bulk materials, such as RuO$_2$~\cite{berlijn2017itinerant,zhu2019anomalous,zhou2024crystal,fedchenko2024observation}, MnTe~\cite{gonzalez2023spontaneous,krempasky2024altermagnetic}, MnTe$_2$~\cite{zhu2024observation}, and CrSb~\cite{li2025topological,lu2025signature,ding2024large,zhou2025manipulation,yang2025three}, Rb$_{1-\delta}$V$_2$Te$_2$O~\cite{zhang2024crystal}, and KV$_2$Se$_2$O~\cite{jiang2025metallic}. However, 2D example has not been demonstrated in experiment so far. In view of 2D materials' high flexibility, tunability, and potential for nanoscale device applications,
it is much desired to find robust 2D altermagnetic materials, preferably also having other emergent degrees of freedom.

In this work, through first-principles calculations and theoretical analysis, we show that the family of monolayer iron oxyhalides
Fe$_2X_2$O ($X=$Cl, Br, I) is a new class of 2D altermagnetic RCIs. These materials exhibit excellent stability and robust
$d$-wave altermagnetism.  They are semiconductors, each of their spin channels possesses a nontrivial real Chern number
($\nu_R^\uparrow = \nu_R^\downarrow = 1$), so the whole system can be viewed as a combination of two copies of spin-polarized
RCIs. This real Chern topology, coupled with altermagnetism, gives rise to spin-polarized topological corner modes. Specifically, in a $C_4$-symmetric sample, corner modes with opposite spin polarizations reside at different corners and are connected by fourfold rotation. Moreover, these 2D altermagnets host an intriguing valley structure, demonstrating a strong coupling between altermagnetic order, valley, spin, and lattice. We predict a valley linear dichroism, enabling selective excitation of altermagnetic valleys
by linearly polarized light, and the excited valley-polarized carriers are also simultaneously spin-polarized.
Applying lattice strain can also induce valley polarization and split corner modes. Notably, Fe$_2$Cl$_2$O is multiferroic, exhibiting both ferroelasticity and altermagnetism, where a small strain can switch its N\'eel vector orientation. In addition, the charge and spin conductivities of these altermagnets are readily tunable by valley/spin polarization and by strain. Our findings establish monolayer iron oxyhalides as a promising platform for exploring magnetic real Chern topology and its interplay with altermagnetism, spin, valley, and lattice degrees of freedom.

%\section{Results and Discussion}
%\subsection{Crystal structure}
\begin{figure}
	\centering
	\includegraphics[width=0.9\columnwidth]{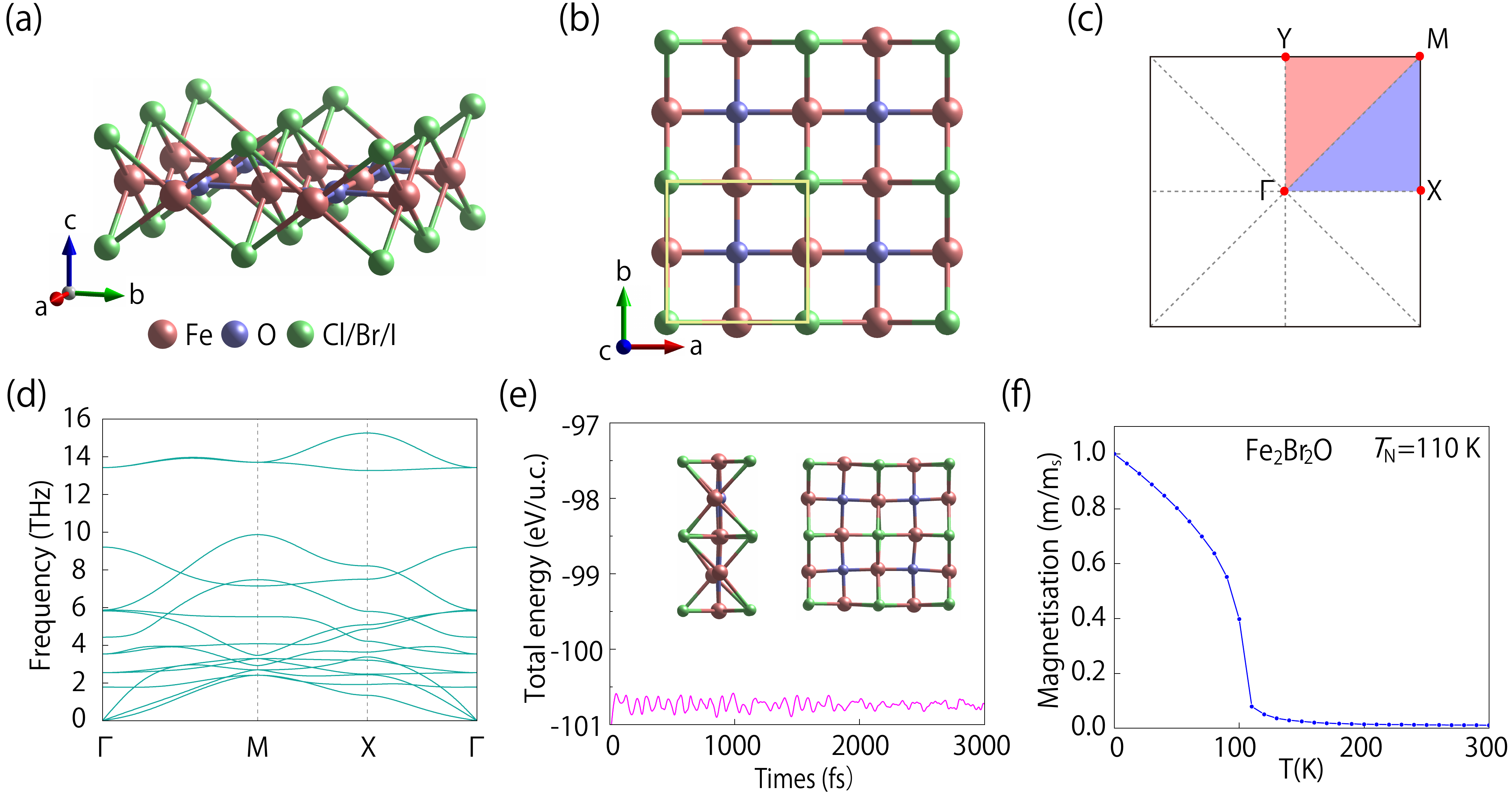}
	\caption{(a) Side view and (b) top view of the crystal structure of monolayer Fe$_2X_2$O ($X$ = Cl, Br, I). (c) Brillouin zone, with high-symmetry points indicated. (d) Calculated phonon spectra of Fe$_2$Br$_2$O. (e) Result of ab-initio molecular dynamics simulations (AIMD) for Fe$_2$Br$_2$O. Panel (f) shows the normalized magnetic moments of monolayer Fe$_2$Br$_2$O as functions of temperature, obtained from Monte Carlo simulations.
		\label{fig1}}
\end{figure}

Our proposed 2D iron oxyhalides Fe$_2X_2$O ($X=$Cl, Br, I) adopt the same crystal structure as V$_2$Se$_2$O and
V$_2$Te$_2$O~\cite{lin2018structure,ablimit2018v2te2o,ma2021multifunctional}. As shown in Fig.~\ref{fig1}(a) and (b), the structure
consists of three atomic layers. The central layer is composed of Fe and O atoms, forming a Lieb lattice. This layer is sandwiched by two outer layers of halide atoms. In the top view (Fig.~\ref{fig1}(b)), the halide atoms are sitting at
the central sites of the O squares. The crystal structure is centrosymmetric, having a space group of $P4/mmm$ (No.~123) and a point group of $D_{4h}$.

The optimized lattice parameters and cohesive energies of monolayer Fe$_2X_2$O ($X =$ Cl, Br, I) are listed in Table~\ref{table1}. All calculated cohesive energies exceed 19 eV per formula unit, indicating strong bonding. The phonon spectra (Fig.~\ref{fig1}(d) and Fig.~S1) show no imaginary modes, confirming dynamical stability. \emph{Ab-initio} molecular dynamics simulations further demonstrate that the structures remain robust up to 300 K (Fig.~\ref{fig1}(e) and Fig.~S1), evidencing excellent thermal stability.
\begin{table}[htb]
	\centering
	\caption{\label{table1} Calculated results for Fe$_2X_2$O ($X $= Cl, Br, I), including optimized lattice constant $a$, cohesive energy $E_\text{c}$, Néel temperature $T_{N}$, and band gaps obtained using PBE ($E_\text{PBE}$) and HSE06 ($E_\text{HSE}$) methods.}
	\begin{tabular}{ccccccccc}
		\hline\hline
		System  & $a$ (\AA) & $E_\text{c}$ (eV) & $T_{N}$ (K) & $E_\text{PBE}$ (eV) & $E_\text{HSE}$ (eV) \\
		\hline 	
		{Fe\(_2\)Cl\(_2\)O}  & 4.035 & 22.10  &120  & 2.924 & 3.69 \\			
		%\hline 	
		{Fe\(_2\)Br\(_2\)O}  &  4.067 & 20.97 &110  & 2.765 & 3.67  \\		
		%\hline 				
		{Fe\(_2\)I\(_2\)O}  &  4.123 &  19.64 &97   & 2.386 & 3.20  \\
		\hline\hline
	\end{tabular}
\end{table}

%\subsection{Altermagnetic ordering}\label{AO}
Iron, as a $3d$ transition metal element, often introduces magnetism. Our calculation shows that the usual checkerboard AFM configuration  is favored for all three members of the family (Fig.~S2).
Importantly, this AFM ordering makes these materials belong to the class of $d$-wave altermagnets.
This is because the magnetism preserves the inversion symmetry $\mathcal{P}$, but breaks the $\mathcal{PT}$ symmetry, where $\mathcal{T}$ is the time reversal operation. The two magnetic sublattices cannot be connected by inversion or translation, instead, they are connected by a fourfold rotation, i.e.,
the system has  $C_{4z}\mathcal{T}$ symmetry. This characterizes these materials as $d$-wave altermagnets. In the absence of SOC, the magnetic state belongs to spin space group No.~47.123.1.1~\cite{liu2022spin,chen2024enumeration}.

To determine the magnetic anisotropy, we perform calculations with SOC included. We find that both
Fe$_2$Br$_2$O and Fe$_2$I$_2$O favor a N\'eel vector in the out-of-plane ($z$) direction, whereas Fe$_2$Cl$_2$O prefers an in-plane direction along the crystal axis ($x$ or $y$ direction). It follows that the magnetic ordering
in Fe$_2$Cl$_2$O spontaneously breaks its in-plane isotropy, leading to ferroelasticity and multiferroic behavior (see Supporting Information).

We estimate the magnetic transition temperatures by using Monte Carlo simulations. As shown in Fig.~\ref{fig1}(f) and Fig.~S1, the estimated transition temperatures are around 100 K. The specific values are presented in Table~\ref{table1}. Although these values are below room temperature, they are still higher than those for many experimentally reported 2D magnetic materials, such as CrI$_3$~\cite{huang2017layer,jiang2018controlling} and Fe$_3$GeTe$_2$~\cite{deng2018gate,fei2018two}.

%\subsection{Band structure and valley structure}

\begin{figure}
	\centering
	\includegraphics[width=0.95\columnwidth]{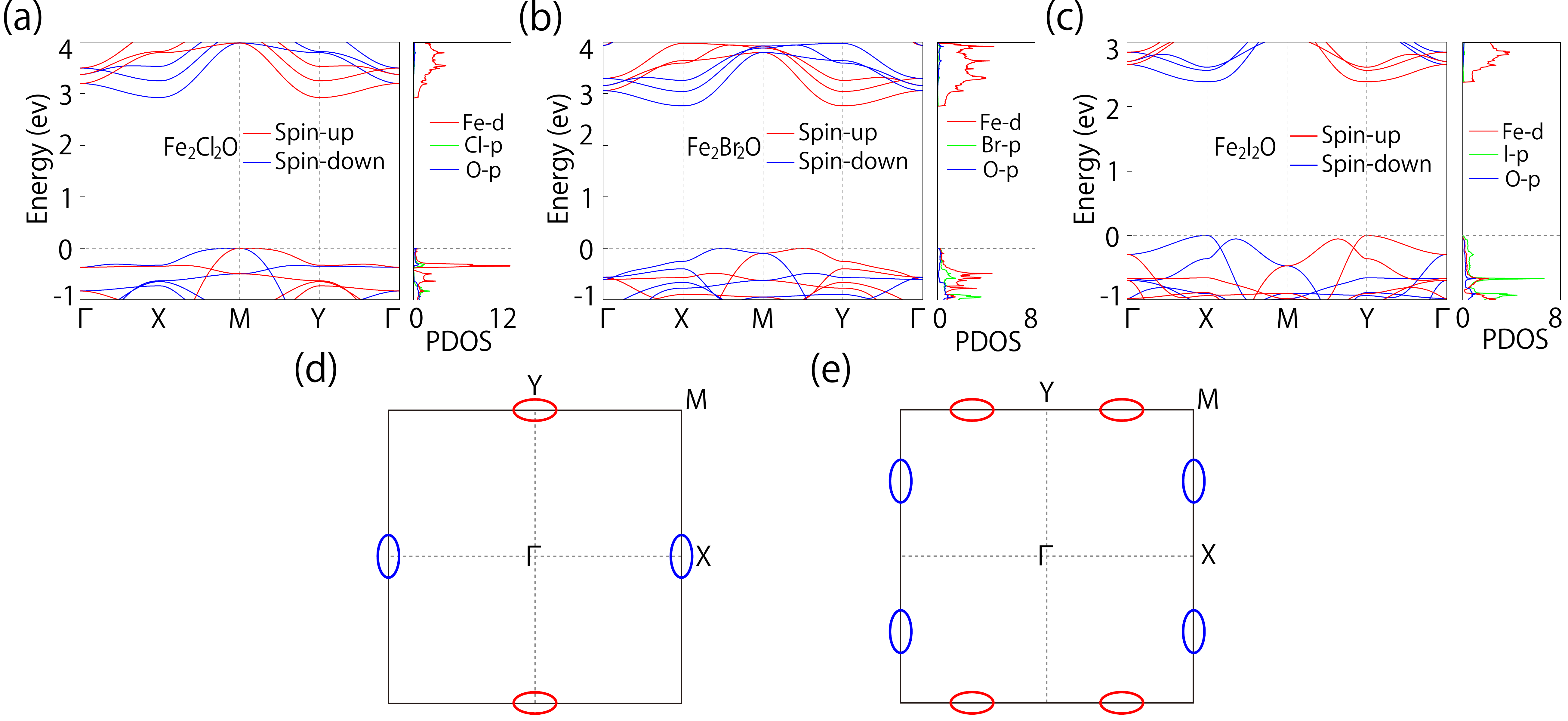}
	\caption{ Panels (a)–(c) show the band structures and projected density of states (PDOS) of monolayer Fe$_2$Cl$_2$O, Fe$_2$Br$_2$O, and Fe$_2$I$_2$O, respectively. SOC is neglected in the calculation. Red and blue colors denote spin-up and spin-down states, respectively. Panel (d) illustrates the two conduction band valleys of these materials. Panel (e) depicts the four valence band valleys of monolayer Fe$_2$Br$_2$O. Red and blue denote spin-up and spin-down polarizations, respectively.
		\label{fig2}}
\end{figure}

Next, we investigate the electronic band structures of monolayer Fe$_2X_2$O ($X=$Cl, Br, I) in their altermagnetic ground states.
In Fig.~\ref{fig2}(a)-(c), we plot the calculated band structures in the absence of SOC. (We find SOC has only weak effects on the low-energy band structure (see Fig.~S3).)
Here, the bands of spin-up (spin-down) channel are marked with red (blue) color. One observes that the spin-up and spin-down bands split in energy-momentum space.
Moreover, the spin splitting exhibits a $d$-wave pattern, e.g., the splitting has an opposite sign between $\Gamma$-$X$ and $\Gamma$-$Y$ paths, ensured by $C_{4z}\mathcal T$ symmetry. This type of spin splitting in the absence of SOC is the hallmark of altermagnetism.

All three iron oxyhalides are insulating with sizable band gaps exceeding 2 eV (Table~\ref{table1}). Band structure results using the hybrid functional (HSE06) are shown in Fig.~S4, and the
corresponding gap values are listed in Table~\ref{table1}. From the projected density of states (PDOS), one can see that for Fe$_2$Cl$_2$O and Fe$_2$Br$_2$O, the conduction and valence band edges are dominated by Fe $3d$ orbitals. As for Fe$_2$I$_2$O, the I-$5p$ orbitals make more contribution at the valence band maximum (VBM).

Interestingly, these materials offer promising valleytronic platforms with spin-polarized valleys. In  Fe$_2$Cl$_2$O, it has an indirect band gap, with CBM located at a pair of energy-degenerate valleys at $X$ and $Y$ points (see Figs.~\ref{fig2}(a) and ~\ref{fig2}(d)), and with VBM at $M$ point.
Fe$_2$Br$_2$O has similar feature as Fe$_2$Cl$_2$O in terms of indirect gap and the conduction band valley structure, however, its VBM is not at $M$ but at some point along $M$-$X$ (and $M$-$Y$) path. This leads to four valleys for its valence band, as shown in Figs.~\ref{fig2}(b) and~\ref{fig2}(e). Different from the other two, Fe$_2$I$_2$O has direct band gap, and both conduction and valence bands exhibit two valleys at $X$ and $Y$ points. It is important to note that these valleys are spin polarized, exhibiting a spin-valley coupling. For example, considering their conduction band valleys in Fig.~\ref{fig2}(d), the $X$ valley is spin-down, whereas the $Y$ valley is spin-up. Their opposite spin polarizations are dictated by
$C_{4z}\mathcal T$ symmetry. This strong spin-valley coupling can be utilized to manipulate spin and valley degrees of freedom.

%\subsection{Real Chern topology and corner modes}

In an altermagnet, the two spin channels are decoupled in the absence of SOC. And each spin channel can be effectively regarded as a spinless system, which possesses an effective time reversal symmetry $\mathcal T'$~\cite{vanderbilt2018berry,wu2018nodal}. It follows that each spin channel becomes real, i.e., having real wave functions, due to the symmetry $\mathcal{PT'}$, and we can investigate their real Chern topology.

A RCI is characterized by a nontrivial real Chern number $\nu_R$. For a system with preserved inversion symmetry $\mathcal P$,  $\nu_R$ can be determined from the parity eigenvalues of the valence states
at the four inversion-invariant points $\Gamma_i$ ($i=1,2,3,4$) of the Brillouin zone, according to the following formula~\cite{ahn2019stiefel,chen2021graphyne}:
\begin{eqnarray}
	(-1)^{\nu_R} = \prod_i (-1)^{\lfloor n_-^{(\Gamma_i)} / 2 \rfloor}, \label{RCM}
\end{eqnarray}
where \( \lfloor \cdots \rfloor \) denotes the floor function, 
and \( n_-^{(\Gamma_i)} \) is the number of occupied states at point \( \Gamma_i \) with negative parity eigenvalues.
A nontrivial topological phase is indicated by \( \nu_R = 1 \).
\begin{table*}[htb]
	\centering
	\caption{\label{table2} Parity information at the four inversion-invariant points of monolayer Fe$_2X_2$O ($X$ = Cl, Br, I). The points are $\Gamma$ (0, 0), X (0.5, 0), Y (0, 0.5), and M (0.5, 0.5). Here, $n_+$ ($n_-$) denotes the number of occupied bands with positive (negative) parity eigenvalues. We find the real Chern numbers $\nu_{R}^{\uparrow}$ and $\nu_{R}^{\downarrow}$ are both equal to 1.}
	\begin{tabular}{cccccccccc}
		\hline
		\multicolumn{6}{c}{Spin-Up} &
		\multicolumn{4}{c}{Spin-Down} \\
		\cline{2-5}
		\cline{7-10}
		& $\Gamma$ & X & Y & M & & $\Gamma$ & X & Y & M \\
		\hline
		$n_+$ & 11 & 8 & 12 & 5 & & 11 & 12 & 8 & 5 \\
		$n_-$ & 7 & 10 & 6 & 13 & & 7 & 6 & 10 & 13 \\
		\hline
		\multicolumn{6}{c}{$\nu_{R}^{\uparrow} = 1$} &
		\multicolumn{4}{c}{$\nu_{R}^{\downarrow} = 1$} \\	
		\hline
	\end{tabular}	
\end{table*}

Now, for an altermagnetic insulator without SOC, a real Chern number can be evaluated for each of its two spin channels. Hence, the system is characterized by a pair of real Chern numbers $(\nu_R^\uparrow, \nu_R^\downarrow)$. Nevertheless, here, the two spin channels are not independent, but symmetry connected. In monolayer Fe$_2X_2$O ($X=$Cl, Br, I), they are connected by $C_{4z}\mathcal T$, dictating that
\begin{equation}
	\nu_R^\uparrow=\nu_R^\downarrow,
\end{equation}
namely, the two channels must have the same topological character.

We analyze the parity eigenvalues of the first-principles band structures for monolayer Fe$_2X_2$O ($X=$Cl, Br, I), and find that they share the same result, as shown in
Table~\ref{table2}. Using formula (\ref{RCM}), one finds that $\nu_R^\uparrow=\nu_R^\downarrow=1$, i.e., both spin channels are topologically nontrivial. Thus, monolayer Fe$_2X_2$O ($X=$Cl, Br, I) can be regarded as consisting of two copies of RCIs, one for each spin channel. This confirms that monolayer Fe$_2X_2$O ($X=$Cl, Br, I) are indeed altermagnetic RCIs.

\begin{figure}
	\centering
	\includegraphics[width=0.95\columnwidth]{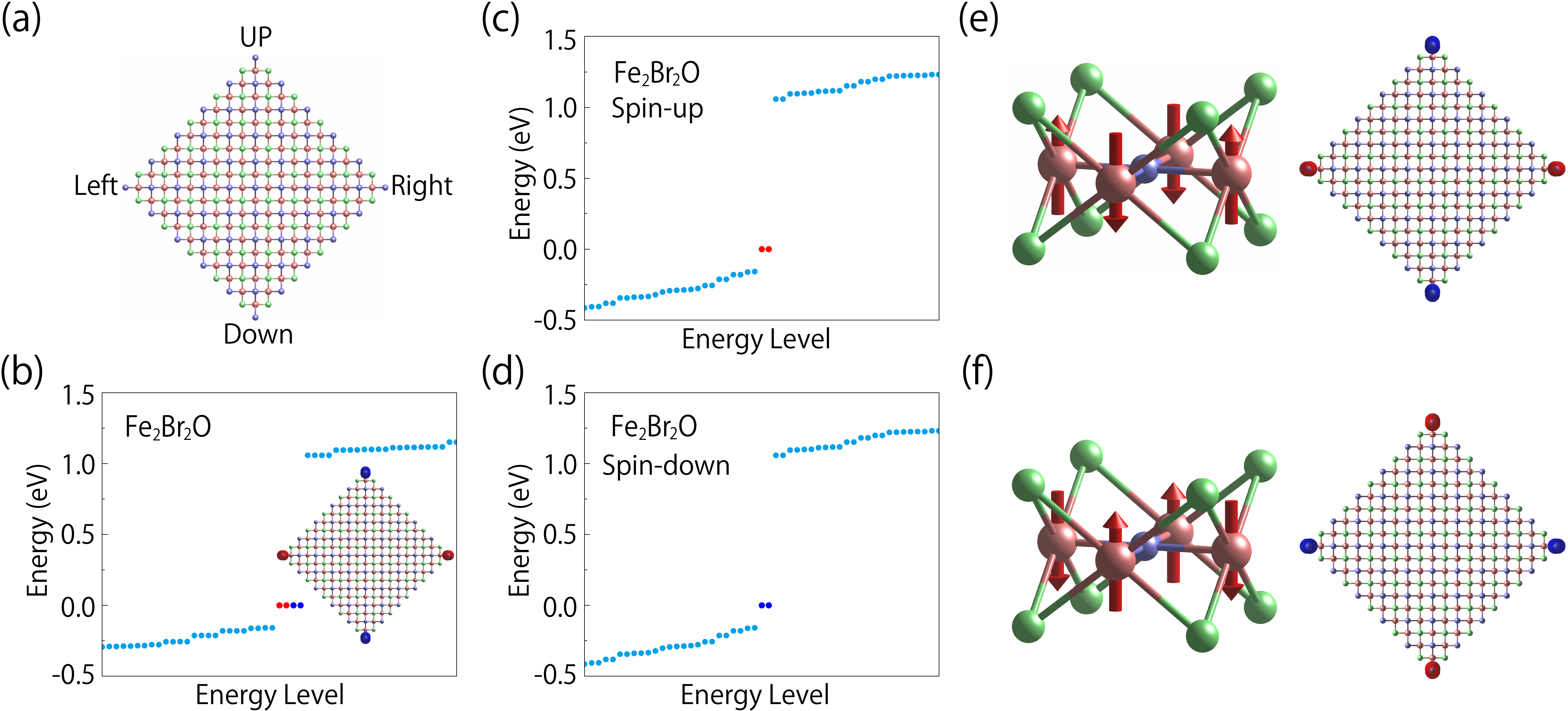}	
	\caption{ (a) The nanodisk geometry used to calculate the corner states. (b) Energy spectrum of the  Fe$_2$Br$_2$O nanodisk, showing a group of four isolated corner modes within the bulk gap. Insets show the charge density distributions of the two groups of corner modes. (c) and (d) show the energy spectra for the spin-up and spin-down channels, separately. (e) and (f) demonstrate that the spin polarization of the corner modes is tied to the AFM N\'eel vector. The spin polarization is reversed upon flipping the N\'eel vector.
		\label{fig3}}
\end{figure}

RCIs have zero-energy modes localized at its corners. It has been shown that in any sample geometry that preserves $\mathcal{P}$ symmetry, the corner modes necessarily come in $\mathcal{P}$-related pairs~\cite{chen2021graphyne}. In an altermagnetic RCI, there is an additional character: The corner modes with opposite spin polarizations are also connected by the
altermagnetic symmetry that link the two spin channels. For monolayer Fe$_2X_2$O ($X=$Cl, Br, I), this symmetry is $C_{4z}\mathcal{T}$. To see this, we take a nanodisk geometry as shown in Fig.~\ref{fig3}(a), which preserves
the fourfold rotation. In Fig.~\ref{fig3}(b), taking Fe$_2$Br$_2$O as an example, we plot its calculated energy spectrum in the nanodisk geometry. One can observe four zero-modes in the energy gap, separated from the other states. These zero-modes are located at the Fermi energy (set as zero here). Moreover, by examining their wave functions, one can see that they are localized at
the four corners, and they are spin-polarized, as illustrated in Fig.~\ref{fig3}(b). To see this feature more clearly, in Fig.~\ref{fig3}(c) and (d),
we plot the spectra of the two spin channels separately. One finds that although the two channels' spectra look identical,
the spatial distribution of their zero-modes are very different. The two spin-up zero-modes are located at the left and the right corners; whereas the two spin-down modes are
at the top and the bottom corners. The two modes in the same spin channel are connected by $\mathcal P$, and the two channels are connected by $C_{4z}\mathcal{T}$, forming this special arrangement. This differs from that in previously studied ferromagnetic RCIs~\cite{chen2020universal,li2022robust}, where the corner modes belong to the same spin channel. The spin pattern of corner modes here can be reversed by flipping the N\'eel vector direction, as illustrated in Fig.~\ref{fig3}(e) and (f). In experiment, the spin-polarized corner modes can be probed by the spin-polarized scanning tunneling spectroscopy technique.

%\subsection{Altermagnetic valleys and valley linear dichroism}{\label{soc}}

In discussing Fig.~\ref{fig2}, we have noted that monolayer altermagnetic Fe$_2X_2$O ($X =$Cl, Br, I) possess spin-polarized valleys in their low-energy band structures. This  is a desired feature for a valleytronic platform~\cite{schaibley2016valleytronics,vitale2018valleytronics}. A key component in valleytronics is to control carriers' valley polarization. This can be easily achieved by applying a magnetic field or coupling to a magnetic substrate~\cite{cai2013magnetic,zhang2016large}. In this case, the two valleys naturally split in energy, creating valley polarization for the electron carriers. Another way is through optical pumping of carriers. We find that for monolayer Fe$_2$Cl$_2$O and Fe$_2$Br$_2$O, there is an interesting effect of valley linear dichroism. This means that by controlling the polarization of incident linearly-polarized light, one can selectively excite carriers that are polarized in one of the valleys. Take Fe$_2$Br$_2$O as an example. We calculate the $k$-resolved linear dichroism parameter $\eta$, defined by
\begin{equation}
	\eta(\bm k) = \frac{|\mathcal{M}_x(\bm k)|^2-|\mathcal{M}_y(\bm k)|^2}{|\mathcal{M}_x(\bm k)|^2+|\mathcal{M}_y(\bm k)|^2},
\end{equation}
where $\mathcal M_i=\langle u_{c}(\bm k)|\partial_{k_i}\mathcal H|u_v(\bm k)\rangle$ ($i=x,y$) is the coupling strength of
interband optical transition with $\sigma_i$ linearly-polarized light, $|u_{c/v}\rangle$ is the cell-periodic part of Bloch wave function of conduction/valence band, and $\mathcal H$ is the Bloch Hamiltonian. For Fe$_2$Br$_2$O, its conduction and valence band states have symmetry characters belonging to $\Gamma_3^{-}$ and $\Gamma_4^{+}$ representations of the $C_{2h}$ point group at $X$ ($Y$) point.
\begin{figure}
	\centering
	\includegraphics[width=0.65\columnwidth]{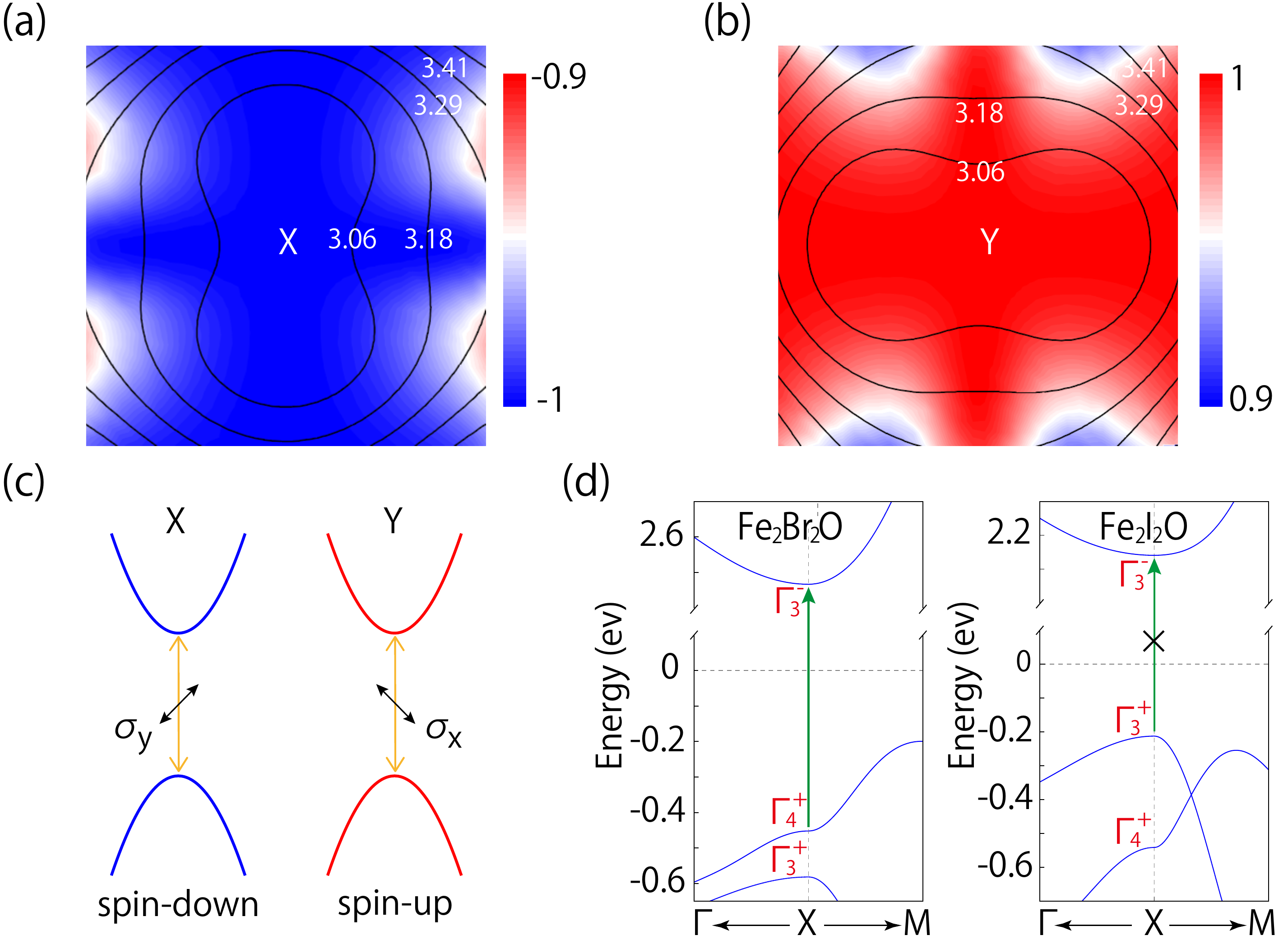}
	\caption{ (a,b) Calculated linear dichroism $\eta(\bm{k})$ near the (a) $X$ and (b) $Y$ points of monolayer Fe$_2$Br$_2$O. The lines indicate the equi-energy contours of the local band gap.  (c) Schematic illustration of the optical transition selection rules for the two valleys. (d) The enlarged band structures around the $X$ point of monolayer Fe$_2$Br$_2$O and Fe$_2$I$_2$O, where the irreducible representations of the top two valence bands and the lowest conduction band are indicated. One can see that in Fe$_2$Br$_2$O, the two states have $\Gamma_4^{+}$ and $\Gamma_3^{+}$ characters, whereas in Fe$_2$I$_2$O, their order is reversed.
		\label{fig4}}
\end{figure}

In Figs.~\ref{fig4}(a) and~\ref{fig4}(b) , we plot the DFT result (with SOC included) of $\eta$
in the regions around $X$ and $Y$ valleys. One observes that
at $X$ valley, $\eta$ is negative, with values close to $-1$.  Meanwhile, $\eta$ at $Y$ valley has an opposite sign and is close to $+1$. The opposite $\eta$ at the two valleys is a feature dictated by the altermagnetic $C_{4z}\mathcal T$ symmetry.
This result shows that the transition at $X$ ($Y$) valley is predominately coupled to $\sigma_y$ ($\sigma_x$) polarized light, exhibiting a strong valley linear dichroism, as illustrated in Fig.~\ref{fig4}(c). Thus, by optical pumping with $\sigma_x$ ($\sigma_y$) polarized light, one can selectively excite carriers in the $Y$ ($X$) valley, creating valley-polarized carriers. Furthermore, since spin and valley degrees of freedom are strongly coupled in these materials,
the valley polarized carriers are also spin polarized. Such valley/spin polarized carriers generated by optical dichroism will provide rich opportunity for valleytronic/spintronic applications.

For Fe$_2$I$_2$O, the situation differs from the above. The optical transition between the VBM and CBM at the $X$ ($Y$) point is forbidden because the two states share the same parity. Comparing the band structures of Fe$_2$I$_2$O and Fe$_2$Br$_2$O (Fig.~\ref{fig4}(d)) reveals that this arises from a band inversion of the top two valence bands. Consequently, although Fe$_2$I$_2$O has a direct band gap, optical transitions between its VBM and CBM are strongly suppressed. Instead, the linear dichroism will appear for light with a frequency larger than the band gap, between the lower valence band ($\Gamma_4^{+}$) and CBM.

Experimentally, the valley linear dichroism in these materials can be detected by first exciting the system 
with linearly polarized light and then measuring the degree of valley polarization through the polarization of the resulting photoluminescence~\cite{zeng2012valley,mak2012control}. In addition, because the two valleys each has a large in-plane anisotropy, the optically excited valley polarization may also be detected via the anisotropy in resistivity. 

%%353
%\subsection{Strain control of valley/spin and multiferroic coupling}

%\subsection{Spin and valley transport}

The transport properties of monolayer iron oxyhalides exhibit interesting features arising from altermagnetism.  In the absence of SOC, spin transport is characterized by two conductivity components, $\sigma_{xx}^\uparrow$ and $\sigma_{xx}^\downarrow$, where the $C_{4z}\mathcal{T}$ symmetry ensures $\sigma_{yy}^\uparrow=\sigma_{xx}^\downarrow$ and $\sigma_{yy}^\downarrow=\sigma_{xx}^\uparrow$.
For an applied electric field $\bm E=E(\cos\theta,\sin\theta)$ along an in-plane direction, where $\theta$ is the angle measured from $x$ axis, the resulting current $\bm j$ has a longitudinal component along $\bm E$ and a transverse component along $\hat z\times\bm E$. Accordingly, one may define
longitudinal and transverse conductivities $\sigma_L^{\uparrow(\downarrow)}$ and $\sigma_T^{\uparrow(\downarrow)}$ for the spin-up (spin-down) channel, such that
\begin{equation}
	j_L^\sigma=\sigma_L^\sigma E,\qquad j_T^\sigma=\sigma_T^\sigma E,
\end{equation}
where $\sigma=\uparrow,\downarrow$.
The charge current $\bm j$ and the spin current $\bm j^s$ can be defined as
\begin{equation}
	j_\alpha\equiv j_\alpha^\uparrow+ j_\alpha^\downarrow,\qquad j^s_\alpha\equiv j_\alpha^\uparrow- j_\alpha^\downarrow,
\end{equation}
where $\alpha=L,T$. Their ratio with $E$ defines the charge conductivity $\sigma_\alpha(\theta)$ and
spin conductivity $\sigma^s_\alpha(\theta)$. A straightforward calculation shows that
\begin{equation}
	\sigma_L=\sigma_{xx}^\uparrow+\sigma_{xx}^\downarrow,\qquad
	\sigma_T=0,
\end{equation}
and
\begin{equation}
	\sigma_L^s(\theta)=(\sigma_{xx}^\uparrow-\sigma_{xx}^\downarrow)\cos 2\theta,\qquad
	\sigma_T^s(\theta)=-(\sigma_{xx}^\uparrow-\sigma_{xx}^\downarrow)\sin 2\theta.
\end{equation}
The charge conductivity is isotropic and has no transverse component. Meanwhile, the spin conductivity has both longitudinal and transverse components, both involving $(\sigma_{xx}^\uparrow-\sigma_{xx}^\downarrow)$,  and their $\theta$ dependence shows a $\pi$ periodicity.
The maximal longitudinal spin current $j_L^s$ is achieved when $E$ is along the crystal axis. And the maximal
transverse spin current $j_T^s$ occurs when $E$ is along the diagonal $\langle 11\rangle$ directions.
\begin{figure}
	\centering
	\includegraphics[width=0.95\columnwidth]{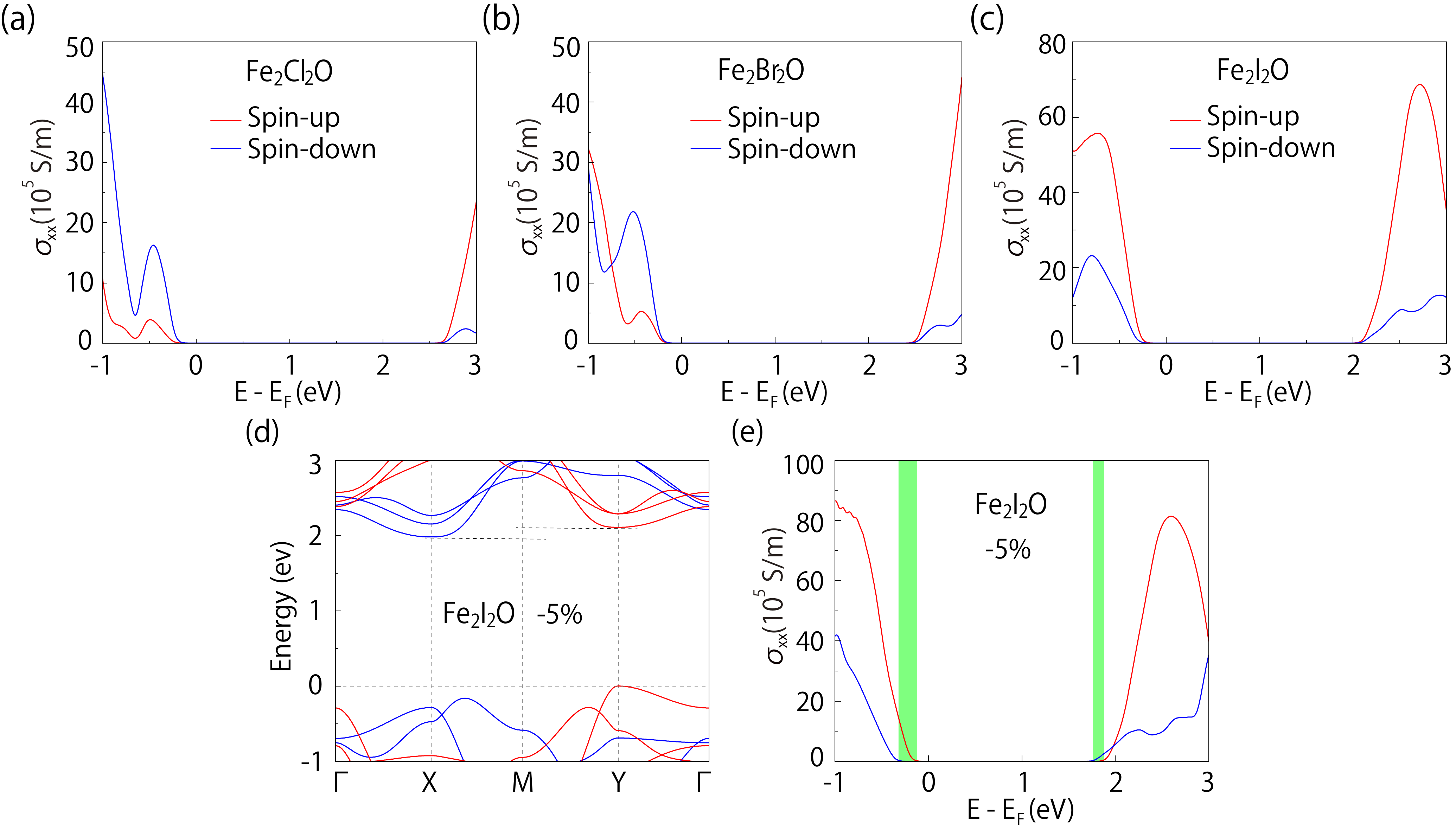}
	\caption{ Spin-resolved longitudinal conductivity $\sigma_{xx}$ of  (a) Fe$_2$Cl$_2$O, (b) Fe$_2$Br$_2$O, and (c) Fe$_2$I$_2$O.
		(d) Band structure of Fe$_2$I$_2$O under -5\% strain, and (e) its spin-resolved charge conductivity. The green color marks the regions where the current is fully spin-polarized.
		\label{fig5}}
\end{figure}

In Fig.~\ref{fig5}, we plot the computed $\sigma_{xx}^\uparrow$ and
$\sigma_{xx}^\downarrow$ as a function of chemical potential for the three 2D iron oxyhalides, using the Drude formula. One can see that for all three materials, the two components exhibit a large difference on the $n$-doped side. This is due to the relatively large valley anisotropy for the conduction band valleys, which can be seen from the anisotropic shape of equi-energy contours in Fig.~S6.

In an altermagnet, the two spin channels are symmetry connected, so generally one cannot have a fully polarized transport by making one of $\sigma_{xx}^\uparrow$ and $\sigma_{xx}^\downarrow$ vanish. Nevertheless, this can be achieved by breaking the $C_{4z}\mathcal T$ symmetry.
For example, by applied magnetic field, magnetic proximity coupling, optical pumping with linearly-polarized light, and applied strain, one can create spin-polarized carriers in one of the valleys. In Fig.~\ref{fig5}(e), we plot $\sigma_{xx}^\uparrow$ and
$\sigma_{xx}^\downarrow$ calculated for Fe$_2$I$_2$O under a uniaxial strain $-5\%$. One observes that there are energy windows in which
one of the spin channels is absent. This can be traced to the band structure in Fig.~\ref{fig5}(d), noting that there is only one spin channel in these windows. It follows that the charge current there will be fully spin-polarized. For doping level close to CBM (or VBM), the current is fully valley-polarized as well.

Besides affecting transport, uniaxial strain that breaks the $C_{4z}\mathcal T$ symmetry can also induce valley polarization and split the energies of spin-up and spin-down corner modes. Moreover, with in-plane N\'eel vector,
we find that Fe$_2$Cl$_2$O is a multiferroic, with both magnetic ordering and ferroelasticity. Uniaxial strain can 
be used to switch its N\'eel vector in the plane, as discussed in Supporting Information.

%\section{Conclusion}
In summary, our study establishes monolayer iron oxyhalides Fe$_2X_2$O ($X$ = Cl, Br, I) as a novel family of 2D altermagnetic real Chern insulators, characterized by nontrivial real Chern numbers and spin-polarized topological corner modes in each spin channel. The strong altermagnetism-valley-spin-lattice interactions in these materials lead to innovative phenomena, such as valley linear dichroism, strain-induced valley polarization, tunable charge/spin conductivities, and multiferroic coupling enabled switching of Néel vector. These findings highlight the profound potential of iron oxyhalides as versatile platforms for advancing spintronics, valleytronics, and quantum technologies, bridging magnetic topology with emergent degrees of freedom in 2D materials.

\begin{suppinfo}	
		The Supporting Information contains: computation methods, stability and magnetic transition temperatures of Fe$_2$Cl$_2$O and Fe$_2$I$_2$O, magnetic configurations, band structures with SOC and permittivity, band structures from the hybrid functional approach, energy spectra and corner modes of Fe$_2$Cl$_2$O and Fe$_2$I$_2$O nanodisks, local band gaps and linear dichroism of Fe$_2$Cl$_2$O, strain control of valley/spin and multiferroic coupling, and band structure results with different $U$ values.		
\end{suppinfo}	

	\begin{acknowledgement}
The authors thank D. L. Deng for helpful discussions. This work is supported by the National Natural Science Foundation of China (Grant No. 12204378), the Key Program of the Natural Science Basic Research Plan of Shaanxi Province (Grant No. 2025JC-QYCX-007), and the Hong Kong Polytechnic University start-up grant (P0057929).
	\end{acknowledgement}

%\bibliography{ref}

\providecommand{\latin}[1]{#1}
\makeatletter
\providecommand{\doi}
{\begingroup\let\do\@makeother\dospecials
	\catcode`\{=1 \catcode`\}=2 \doi@aux}
\providecommand{\doi@aux}[1]{\endgroup\texttt{#1}}
\makeatother
\providecommand*\mcitethebibliography{\thebibliography}
\csname @ifundefined\endcsname{endmcitethebibliography}
{\let\endmcitethebibliography\endthebibliography}{}

\end{document}